# Efficient Hybrid Execution of C++ Applications using Intel® Xeon Phi™ Coprocessor

Jiri Dokulil, Enes Bajrovic, Siegfried Benkner, Sabri Pllana, Martin Sandrieser
*Research Group Scientific Computing*
*University of Vienna*
*Austria*
Email: firstname.lastname@univie.ac.at

Beverly Bachmayer
*Software and Solutions Group*
*Intel GmbH*
*Germany*
Email: bev.bachmayer@intel.com

*Abstract*—The introduction of Intel® Xeon Phi™ coprocessors opened up new possibilities in development of highly parallel applications. The familiarity and flexibility of the architecture together with compiler support integrated into the Intel C++ Composer XE allows the developers to use familiar programming paradigms and techniques, which are usually not suitable for other accelerated systems. It is now easy to use complex C++ template-heavy codes on the coprocessor, including for example the Intel Threading Building Blocks (TBB) parallelization library. These techniques are not only possible, but usually efficient as well, since host and coprocessor are of the same architectural family, making optimization techniques designed for the Xeon CPU also beneficial on Xeon Phi. As a result, highly optimized Xeon codes (like the TBB library) work well on both.

In this paper we present a new parallel library construct, which makes it easy to apply a function to every member of an array in parallel, dynamically distributing the work between the host CPUs and one or more coprocessor cards. We describe the associated runtime support and use a physical simulation example to demonstrate that our library construct can be used to quickly create a C++ application that will significantly benefit from hybrid execution, simultaneously exploiting CPU cores and coprocessor cores. Experimental results show that one optimized source code is sufficient to make the host and the coprocessors run efficiently.

## I. INTRODUCTION

The Intel® Xeon Phi™ coprocessor is a new contender in the HPC market. The main idea behind it is to put together a large number of relatively simple cores derived from the well established x86 range. As the name suggests, the core design is based on Intel Architecture processors, but there are important differences between a current Xeon processor and Xeon Phi coprocessor. The number of cores on Xeon Phi coprocessors is 61 cores on one coprocessor, but a single threaded performance of the Xeon Phi cores is lower than with Xeon processors. The Intel Xeon Phi coprocessor comes as a PCI Express card with its own memory. Unlike a GPU, the Xeon Phi coprocessor can be seen as a computer itself – it runs an operating system based on Linux and is capable of working independently from the host computer. Of course, the host has to provide the coprocessor with some services, like power or network access.

The familiarity of the architecture makes it possible to use familiar programming techniques, libraries and even existing source codes. In the ideal case, the only thing necessary to support Xeon Phi coprocessor in an application is to add one switch to the compiler command line. If it is for example a well designed application that uses OpenMP to achieve parallel execution on "traditional" machines, it will work on the Xeon Phi coprocessor as well.

While this makes it easy to use the coprocessor, it prevents us from using the power of the host's CPUs or using multiple Xeon Phi coprocessors. On the whole, it may be a good idea to build an application that is designed to run on the host and that can also use all available resources (the coprocessors) in the system to improve overall performance. In some cases, it is sufficient to statically distribute the work – split it between the host and all of the Xeon Phi coprocessors. But there are some scenarios, where dynamic work allocation is necessary, for instance if there are large and unpredictable differences in the amount of time that is necessary to complete different work items.

We have designed and implemented a C++ library that can be used to build such an application. The library is based on the idea of task parallelism, which has been made popular by the Intel Threading Building Blocks library (TBB) [1]. At the moment, the library supports hybrid for-loop, i.e. it applies a function to every item in a sequence using the host and the Xeon Phi coprocessors. Our library takes care of work distribution, execution control and data transfers, but the actual task scheduling is done by the TBB task scheduler. The work items are sent to the coprocessor in larger blocks and using double buffering to improve throughput and decrease latency.

Our library is completely written in C++, which is necessary, since TBB is also written in C++. Furthermore, Xeon Phi coprocessor applications are built in a different way to other comparable architectures like CUDA or OpenCL, because the source code is compiled for the coprocessor at the same time as the host binary. Also, the compiler has integrated support for making memory transfers and execute functions on the coprocessor. On the whole, the Xeon Phi software stack is easy to use, but very different from the other architectures. So, we decided to build a new application and in doing so demonstrate, that with our library and Xeon Phi coprocessors it is possible to quickly build software that can efficiently run on the host but also achieve significant performance improvements if one or more Xeon Phi coprocessors are present.

To be more specific, we have implemented a popular fluid simulation technique – smoothed particle hydrodynamics (SPH) [2]. It is a relatively simple yet powerful technique that provides nice ratio of data size to computation time. This makes it interesting and suitable for Xeon Phi coprocessors, since data transfers are important to consider due to limited (compared to CPU-to-memory) transfer rate of PCI Express. The results of each simulation step are then rendered into an image. This rendering step is executed in parallel with the following simulation step so they form a two step pipeline. Simulation and rendering use a spatial indexing structure to improve overall performance. Our implementation and optimization work on this structure lead to some interesting observations about the optimization techniques for Xeon Phi, so it is discussed in detail.

The rest of the paper is organized as follows. First, we describe relevant features of the Xeon Phi architecture in Section II. The design of our library is shown in Section III. Section IV briefly introduces the SPH technique. Section V deals with implementation of SPH for the Xeon Phi architecture, most notably the spatial index optimization. Experimental results are provided in Section VI. The paper closes with a discussion of related work, concluding remarks and discussion of future directions.

## II. XEON PHI ECOSYSTEM

Figure 1 depicts the architecture of Xeon Phi coprocessors at a high-level of abstraction. Xeon Phi coprocessors comprise 61 Intel Architecture (IA) cores. Additionally, Xeon Phi coprocessors include memory controllers that support the GDDR5 specification and special function devices such as the PCI Express interface. Xeon Phi cores run independently of each other, and support hardware multi-threading and vector processing. The memory subsytem supports full cache coherence. Cores and other components of Xeon Phi are connected via a ring interconnect. From the software point of view, a Xeon Phi coprocessor is a Symmetric Multi-Processing (SMP) computing domain, which is loosely-coupled to the computing domain of the host.

The Xeon Phi coprocessor is implemented as a PCI Express form-factor add-in card. The high-level software architecture of a system with a host and Xeon Phi coprocessor is depicted in Figure 2, which shows software components relevant to our work. The left-hand side of Figure 2 shows the host software stack that is based on a standard Linux kernel. The software stack for Xeon Phi coprocessor, which is shown on the right-hand side, is based on a modified Linux kernel. The operating system on the Xeon Phi coprocessor is in fact an embedded Linux environment that provides basic functionality such as process creation, scheduling, or memory management.

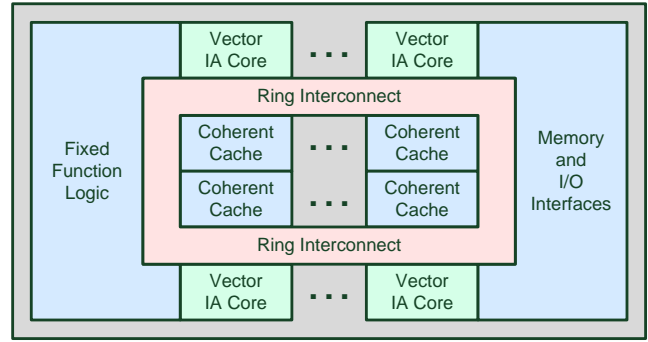

Figure 1. High-level view of the Xeon Phi architecture.

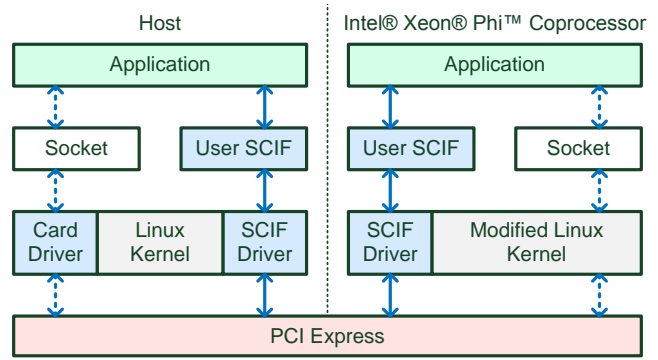

Figure 2. Software architecture of a system with a host and a Xeon Phi coprocessor. Acronym: Symmetric Communication Interface (SCIF).

Multiple options are available for communication between the host and the card. The card driver provides virtual network interfaces, so it is possible to use the TCP/IP network stack. This is good for management and compatibility with existing applications. On the other hand, it cannot provide maximum performance, since the network stack was designed for a different purpose than communication over PCI Express.

The specialized SCIF (Symmetric Communication Interface) library provides two communication options. CPU based messaging interface similar to BSD sockets and DMA transfers. It is a low-level library so there is minimal overhead. Its usage is similar to existing communication libraries for sockets and DMA transfers.

Another important software is the Intel C++ Composer XE compiler. It can do much more than just cross-compile the source for the Xeon Phi platform. There is support integrated into the compiler to build an application that targets both platforms. Effectively, it takes the source code and compiles it for both the host and Xeon Phi coprocessors, optimizing each binary for the specific target architecture. Some parts of the code may be marked (using `#pragma` annotations in the source code) to run on the coprocessor (this is called *offloading*). This means that the marked part of the code needs to be put into a hidden function which is exported by the binary executing on the Xeon Phi coprocessor. The compiler creates the hidden function and also takes care of the remote call of it. It is also possible to transfer various data buffers when the offload is performed, but the details are outside the scope of this description, since we do not use this feature in our application. There is no direct support for hybrid execution – when the offloaded call is made, the calling thread on the host suspends until the offloaded function finishes.

The compiler's offloading support is convenient and very difficult to replicate "by hand" if C++ templates are used. The Intel compiler supports much of the C++11 standard and it is possible to fully use all supported C++ features on the Xeon Phi coprocessor as well, so it may be useful to for example offload calls to function templates. Since the Xeon Phi binary has to export all functions (in this case instances of the function template) that are used in offload calls, it is necessary to figure out every possible template instantiation. Furthermore, C++ name decoration makes it hard to discover correct names of the exports. But the compiler has to discover all template instances anyway and it has no trouble figuring out name decoration, so it can easily provide this functionality automatically.

Let us sum up the ways in which the Xeon Phi card can be used as a coprocessor by a process running on the host. First, one can use the card as if it was another node connected to the same network. The virtual network provided to the card by the host makes it possible. Second, one can use specialized versions of libraries like MKL that use Xeon Phi internally. Third, it is possible to manually perform the same tasks that the compiler does to offload function calls using Xeon Phi libraries. Last but not least, one can use the offloading support provided by the compiler. In this case, the compiler automatically deals with compilation for multiple targets and generates code that automatically deploys the Xeon Phi binary and establishes communication.

The last option (offloading compiler) is the most convenient to use, if the second option (libraries that use Xeon Phi internally) is not appropriate. It is still necessary to use `#pragma`s and their options to specify what data should be transferred and when should the execution be transferred to the coprocessor – this does not happen automagically. This approach integrates well with OpenMP parallelization. However, there is no straightforward way to use these techniques to implement a hybrid application that uses dynamic work allocation. It is possible to combine this approach with the SCIF library to create an application or a library that is still easy to compile and deploy, but more flexible. We have decided to build our library this way.

## III. TBB INSPIRED OFFLOADING LIBRARY

The extensive C++ support and the design principles of TBB gave us the idea to further extend the standard `for_each` function template, which executes a function on each member in a container. The TBB library provides `parallel_for_each` that performs the same thing in parallel. We decided to create the `offload_for_each` function template to perform the same work, but in a hybrid way. We wanted it to execute on the host and on (possibly several) Xeon Phi coprocessors. Furthermore, we didn't want to force the user to make significant modifications to the existing code. It should be possible to swap `for_each` to `offload_for_each` without changing the data representation or the object that represents the action. Instead, the user may be required to define appropriate traits (a trait is a template class that provides type definitions, constants or functions for a data type). The compiler support for C++ templates during offloading is essential for this to work. We are not creating a function but a template function – a function generator that will create a different function for every combination of type parameters, so that all polymorphism is resolved at compile time rather than during run-time.

Since there is no shared memory between the host and the coprocessors, it is not possible to fully mimic the behavior of `for_each`. What `offload_for_each` really does is this: a copy of the function object is created on each of the processors (the host and the coprocessors). Then, the host's copy is executed on some of the items in the input sequence in parallel. The rest of the items are sent (serialized, transferred with DMA and deserialized) to the coprocessors where the local copy of the function object is executed on the items in parallel. The result of the function object's invocation is then sent back to the host and stored back in the original sequence. Each of the items is sent to just one coprocessor, unless it is processed by the host, in which case it is not transferred at all – the function is executed in-place. Note, that items are not sent one item at a time but as larger blocks and also that double buffering is used to reduce latency. The way in which the run-time decides where to execute each item is presented in detail in Section III-C. The execution is synchronized using SCIF messages and the data is transferred using DMA transfers (again, provided by SCIF).

The `offload_for_each` function accepts three parameters. Two random access iterators that define the sequence of items to be processed and one functor (function object) that defines the operation to be performed. The following example shows a way in which the function may be called.

```
std::vector<data_type> data;
fill_data(data);
functor_type functor;
offload_for_each(data.begin(),data.end(),functor);
```

The functor is similar to a traditional C++ function object:

```
struct functor_type
{
  int i;
```

```
  void operator()(data_type& x)
  {
    x=x*i+2;
  }
};
```

The `operator()` performs the actual work. It does so in-place, so it does not return the new, modified version. The effect of the call is similar to one of the following:

```
std::for_each(data.begin(),data.end(),functor);
tbb::parallel_for_each(data.begin(),data.end(),
    functor);
```

In other words, `functor::operator()` is called on each item between `data.begin()` and `data.end()`. But there are some important differences. Each coprocessor has its own copy of the functor, so any changes to members of the functor that are made on the coprocessor are not reflected on the host. Also, if that functor reads or modifies global variables, it does so with their copies on the coprocessor.

### A. Data serialization

Since the Xeon Phi coprocessor is basically a fully functional, separate Linux machine, the first issue to overcome is the fact that the memory of the host is not shared with the coprocessor. This means that all necessary data must be transferred to the Xeon Phi coprocessor and then the result has to be transferred back. However, we do not want to force a specific data representation onto the user, for example requiring that all data is stored as a vector of simple structures. Therefore, we need to provide some mechanism that allows the user of the library to specify which data to transfer and the way the data should be handled during the transfer.

In order to transfer the data to and from the Xeon Phi coprocessor, we have defined a serialization interface. This is one of the situations, where template metaprogramming provided by C++ can be used to make the design clean and easy to use. For every data type that has to be transferred, the user of the library has to specify a type trait that describes the way in which the data type is serialized.

The serialization trait is in fact quite simple. It only needs to provide three functions:

1) serialization of an object,
2) deserialization of an object, and
3) size – report the number of bytes that will be necessary to serialize a concrete variable.

The first two functions are actually function templates. They are parametrized by a reader or writer object. This way, the serialized data can for example be written to a memory buffer or immediately transmitted over a network. Once the data has been serialized to a memory buffer, it can be transferred to the coprocessor using DMA and then deserialized.

Naturally, there are many details that need taking care of in order to serialize very complex structures, especially those that link data with pointers. These may be very tricky to deal with in some cases, but they are outside the scope of this paper. But in simple cases, the definition of seralization is really simple. For example, the following code example is enough to serialize the `functor_type` that was used in the example above.

```
template<> struct serializer<functor_type>
{
  template<typename Writer>
  static void serialize_to_writer(Writer& writer,
      const functor_type& x)
  {
    writer.write(x.i)
  }
  static std::size_t size(const functor_type& x)
  {
    return sizeof(int);
  }
  template<typename Reader>
  static void deserialize(Reader reader,
      functor_type& x)
  {
    reader.read(x);
  }
};
```

### B. Run-time architecture

The TBB library is used not only as an inspiration but also to parallelize execution on the host and on the Xeon Phi coprocessors. This is possible thanks to the specifics of the Xeon Phi architecture and it allows us to move some of the scheduling duties from the host to the individual coprocessors. For efficiency, we are also forced to use operating system threads besides the TBB thread pool. As a result, our run-time environment consists of multiple OS threads and TBB tasks.

Using TBB as scheduler as part of our run-time has several advantages. First, the TBB task scheduler is reliable, very efficient and designed is such a way that it runs well even on the somewhat different architecture of Xeon Phi coprocessor. Second, it allows the users of our library to use TBB in their codes or make multiple concurrent calls to our library without adverse effects, like creating too many worker threads. This is a natural consequence of TBB's design, since the library itself maintains a fixed number of worker threads and the user's code only creates tasks that are not bound to CPU threads.

When the `offload_for_each` function template is called, several preparatory steps have to be performed before the actual computation starts:

- spawn a master thread on each of the coprocessors,
- establish a communication channel to each of the coprocessors,
- transfer (serialize and send) the function object to all of the coprocessors,

- spawn child TBB tasks (workers) on the host that will perform computation on the host, and
- spawn service threads on the host.

The master threads on the coprocessors are quite simple. First, they establish the communication channel to the host, then receive and deserialize the function object. After that, the thread spawns TBB tasks (workers) that take care of the actual computation on the coprocessor. These tasks also deal with the necessary data transfers if necessary. There are two types of the transfers: unprocessed items are being sent from the host and processed items are being sent back to the host. If there is no more work to do, the master thread performs clean-up and terminates.

The workers on the host are even simpler. If a worker has nothing to do, it requests an item to process and then executes the function object on that item. If there are no more items, the worker terminates.

Last but not least, there are the host's service threads. They allocate work to the individual coprocessor cards and deal with all of the communication (including serialization and deserialization) with the cards. In our current design, the following threads are used to serve each coprocessor:

- controller – initializes communication with the card, both direct messaging and DMA, issues work for the other service threads, handles termination and clean-up,
- support worker – performs long (computation or memory heavy) operations for the controller so that controller can quickly respond to any messages without waiting for these operations to complete,
- waiter – waits for the DMA operations to finish and then issues a task to the worker thread.

The waiter thread is necessary, since it is currently not possible to passively wait for SCIF messages and DMA transfers at the same time. Instead, the waiter is given orders to (passively) wait for a DMA transfers to finish and then issues a task. This task may for example be deserialization of a result that has just been transferred from the coprocessor, or it may just send a message to the controller. This way, the controller can wait for messages and still be notified when a DMA transfer finishes.

It would be possible to redesign the service threads so that just one set of threads (controller, support worker and waiter) would be able to control any number of coprocessors. In that case, it may be necessary to increase the number of support worker threads. What we do at the moment is also use the workers that would otherwise perform the computation (execute function object on the data) to act as auxiliary support workers if there are more tasks than the dedicated support worker thread can handle.

Figure 3 shows a slightly simplified version of the start of `offload_for_each` execution. The master thread is the thread from which the function was called. The remaining lifelines show either threads (in the traditional sense) or TBB tasks. The difference is, that a task may or may not be bound to a thread, depending on the task scheduler. The threads are created independently of TBB as operating system threads. In a simple case where just one call to `offload_for_each` is made at one time, the system on the host is running the following threads:

- threads maintained by the TBB task scheduler,
- the master thread,
- the waiter thread, and
- the support worker thread.

If the number of threads maintained by TBB is $N$, there are $N+3$ threads running in the system. By default, $N$ is the number of logical CPU cores available on the host. In general, it is better if the number of active threads does not exceed the number of logical cores, so the operating system is not forced to reschedule the threads. It may look as if our implementation violates this rule, since we have three extra threads. But we need to consider, whether a thread is running or suspended. The master thread is actually suspend just after `offload_for_each` starts and only resumes before the end, so it means we have one active thread less, giving us $N+2$ in total. The waiter thread is also suspended most of the time. It is either (passively) waiting for a DMA transfer to finish or (again passively) waiting for a command from the controller. We are at $N+1$ threads. We cannot discount the support worker thread, since it spends a lot of its time doing work (mosty data serialization and deserialization). But we still get to $N$ active threads, because the controller, which is in fact a TBB root task, is being executed on one of TBB's threads. It has to be, because once it is started, it does not release the thread until it finishes. The controller spends most of its lifetime suspended, (passively) waiting for a message. If it receives one, it either issues work to one of the helper thread or performs a trivial task (mostly buffer allocation and management) and sends a message back. It never does anything computationally heavy, so it can also be seen as a thread which is suspended. In total, we have $N-1$ active TBB threads and several other threads that on average add up to less than one active thread.

The situation on the coprocessor is simpler, since the worker tasks take care of computation, serialization and deserialization. The only work left is to deal with buffer allocation, which is needed only rarely and is trivial. Because Xeon Phi coprocessor provides much larger number of logical CPUs, we can reserve one for this purpose anyway, with very little negative impact on performance.

*C. Work allocation*

The work is allocated to the host and the coprocessors using a priority queue. The workers on the host request work for themselves, the controller requests work on behalf of the coprocessor it controls. All controllers run on the host, so the queue is not distributed.

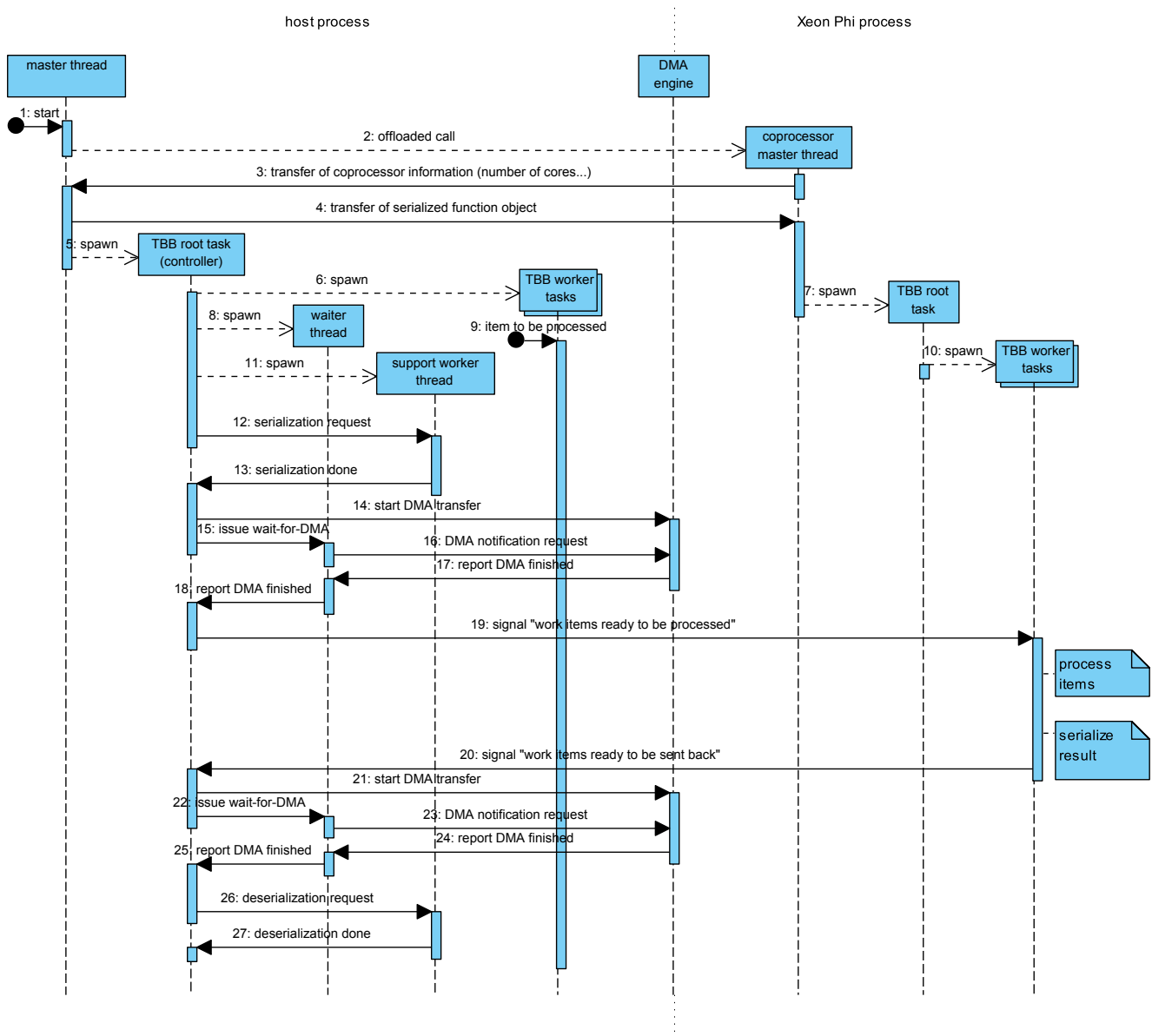

Figure 3. Sequence diagram of initial phase of `offload_for_each` execution. Message 4 may be performed as a DMA transfer if the object is large. The diagram does not display buffer management which is performed by both TBB root tasks and requires them to communicate with each other.

At the beginning, the whole sequence of items passed to the `offload_for_each` is added to the queue with low priority. If any unit (worker or controller) needs an item to process, it dequeues the item at the head of the queue. It is possible to request multiple items. If a unit requests more items than it can process, the excess items are put back in the queue with high priority. This may happen to a controller, since they request items in larger batches. The size of items' serialized form cannot be determined precisely in advance, so the total size may be larger than the available buffer, in which case the excess items are put back in the queue. In our implementation, only the high priority items are actually stored in a queue. A sequence counter is sufficient for the low priority items.

The workers on the host are completely independent. If any of them becomes idle, it tries to get the next available unprocessed item from the queue and processes it. If there are no more items to process, the thread terminates, because all items have either already been processed or allocated to another worker or a Xeon Phi coprocessor.

The controller threads on the host allocate work to their respective Xeon Phi coprocessors. They work independently from each other. To better utilize the DMA subsystem, a controller allocates the work in blocks of multiple items.

The number of items in the block is currently the same as the number of logical cores available on the coprocessor. The items are serialized into one buffer and then send to the coprocessor as one large DMA transfer. As soon as the coprocessor receives the data, the workers (on the coprocessor) can start requesting the items from that block. The controller on the host tries to maintain double buffering. This means that as soon as there is just one block of items present on the coprocessor (including those that have not yet been fully transferred) the host starts preparing the next block and sends it to the coprocessor as soon as possible. This way, there should always be items waiting to be processed on the coprocessor. This is not true near the end of the computation or if the serialization and transfer of the data takes too long compared to the time necessary to process such data – in that case, the host is not able to provide enough data in time. If there is a significant probability of long sequences (comparable to the number of cores on the Xeon Phi coprocessors) of items that are trivial to process, the number of "hot" buffers (those waiting to be processed on the coprocessor) could be increased, providing that the total number of items is large enough to compensate for large scheduling granularity. At the moment, we do not do this in our implementation.

When the processed items are returned to the host, they are also bundled together into one buffer. When that buffer gets full, the last worker to write to it sends it to the host. Whenever possible, the data transfer subsystem on both sides tries to reuse buffers that are no longer needed. For example, if all of the data from a block that has been transferred to the coprocessor has been processed, that block is reused to store the results that will later be sent back to the host. Later, when the results have been transferred back and read (deserialized) from the buffer, the buffer may once again be used to transfer data to the coprocessor.

## IV. SMOOTHED PARTICLE HYDRODYNAMICS

We have chosen smoothed particle hydrodynamics [2] as an example to test our library. This is a very general method for fluid simulation. The idea is to view the liquid as a set of particles that carry the physical properties of the liquid. The position of the particles may arbitrarily change over time. So, it is not a grid based technique. The actual physical properties of the simulated liquid at any point is derived from the physical properties of the particles that lie near that point. The influence of each particle is defined by a kernel function. We use a Monagnan cubic spline [3] as the kernel function. The advantage of this kernel is the fact, that beyond certain radius $h$, the value of the function is 0. So, to evaluate properties of the liquid at any point, we only need to consider particles that are closer than $h$. A spatial index is used to speed up lookup of the particle's neighbors.

We want to simulate a nebula – a gas cloud in space. We are interested in these physical properties:

- mass, that defines inertia,
- pressure, that defines repulsive force, and
- gravity, that defines attractive force.

Both pressure and gravity naturally depend on mass. Unlike pressure, gravity affects particles over larger distances, so we do not simulate gravity interaction between each particle pair. Instead, we simulate a gravity field, that is grid based – each particle contributes to the field in each segment, but the number of segments is limited.

Each simulation step is performed in several phases:

1) Prepare common structures – spatial index and gravity field.
2) Physics simulation part 1 – for each particle, apply gravity field and compute gas density at that particle's position.
3) Physics simulation part 2 – for each particle, evaluate effects of gas pressure on the acceleration of the particle.
4) Clean up – a few trivial operations for every particle (update speed and position).

In general, physical properties $A(x)$ at any position $x$ is given by the following equation:

$$A(x) = \sum_j m_j \frac{A_j}{d_j} W(|x - pos_j|, h)$$

where $p_j$ are the particles, $m_j$ is mass of particle $p_j$, $d_j$ is density associated with $p_j$, $A_j$ is the value of the physical property for particle $p_j$, $|x - pos_j|$ is the distance from the particle $p_j$ to $x$ and $W$ is a kernel function (Monagnan cubic spline in our case). The parameter $h$ is the smoothing radius. The cubic spline is defined like this:

$$\begin{aligned} W(r,h) &= \frac{8}{\pi h^3} * M \\ x &= r/h \\ M &= \begin{cases} 1 - 6x^2 + 6x^3 & \text{if } 0 < x \leq 1/2 \\ 2 * (1-x)^3 & \text{if } 1/2 < x \leq 1 \\ 0 & \text{otherwise} \end{cases} \end{aligned}$$

Since the aim of this paper is to explain our offloading library, we believe it is not necessary to describe the formulas that control the simulation in detail. The simulation is meant purely as an example, so it is not completely physically accurate nor do the simulated scenarios represent any real world data. Unfortunately, none of the existing implementations of SPH we are aware of is written in a way that match the interface of our offloading library, so we had to build a new one.

Still, the SPH problem itself is quite useful as a demonstration in our case. The method is quite simple, the results can easily be interpreted (visualized) and it is not too heavy on the memory. The last aspect is quite important, since the distributed memory architecture does not work well if the data is large but the computational complexity is low. Another nice aspect of SPH is the fact, that it is not possible to accurately predict how much time will be needed to process a particle, because this depends on the number of neighbors of that particle. This allows us to demonstrate flexibility of our library and flexibility of the architecture of Xeon Phi. On the other hand, it is not an ideal case, since it is not sufficient to have data for just one particle in order to perform one simulation step. All particles (or at least the neighbors) are necessary. We always send all particles to the coprocessor as part of the function object (i.e., as one large DMA transfer during startup of the `offload_for_each` call), since it is not possible to quickly determine which particles will be needed. This means, that particles that will be processed on the coprocessor are in fact transferred twice – first time during initialization, second time when they are send to Xeon Phi for processing. This is a price we pay for building the application using the library instead of performing everything by hand. On the other hand, the it is much easier to implement with our library and the experiments show that the price is not that big.

### A. Rendering

We do not just simulate the nebula. The result of each simulation step is rendered into an image. The process (volume rendering) is also not computationally trivial. We have implemented volume ray casting [4]. This is a computationally intense method of rendering volume objects (like gas clouds) that is based on casting a virtual ray from the camera origin through every pixel of the virtual view screen in order to get color of that pixel. As the ray passes through the rendered objects, samples are taken at intervals and the color of the pixel is derived from these samples. This means, that we have to compute physical properties of the gas cloud at each sampled location, so we have to look at all particles that are close enough for the kernel function to be non-zero. The spatial index is once again used to faster find such particles. Note that there are no data dependencies between the pixels, so the algorithm is easy to parallelize.

## V. IMPLEMENTATION OF SPH

In this section, we will discuss our implementation of the SPH technique. Each particle is represented as a simple C++ structure containing its numerical sequential ID, its coordinates in space (x,y,z), its mass, the type of material (used to color the result), current density and pressure at the particle's position, particle's velocity and acceleration. The particles are stored in a contiguous array (to be exact, it is a `std::vector` from C++ Standard Template Library).

The gravity field is represented as a three dimensional matrix of force vectors (force vector is a structure with three members: x, y, z). In memory, the matrix is represented as one contiguous array (again, `std::vector`) to avoid memory fragmentation and speed up serialization.

All values are represented using double-precision floating-point format. We use basic arithmetic operations and square root. The exact number of operations required to process each particle depends on the number of neighbors of that particle that are closer than the smoothing distance $h$. This also means that the total number of operations per simulation step is not constant.

### A. Spatial index optimization

The last structure important for the simulation is the spatial index: we often need to find neighbors of a particle. It would be inefficient to iterate through all particles and test whether the particle is close enough. So, we implemented a spatial index that splits the space into smaller blocks. Then, we only need to evaluate particles in the blocks that are close enough to the particle (or any other point in space). This may be just one block or multiple blocks. We use pre-defined number of blocks and all blocks are the same size. This made the index easy to implement, but there is no guarantee about the number of particles that will be tested unnecessarily. More advanced spatial indexes could be used to that end, but this solution proved sufficient for our purpose.

However, things are not so simple, since the index structure has to be somehow represented in memory. At first, each block was represented by an array (`std::vector`) of pointers to particles that lie in that block. It took a long time to build this index (as much as a quarter of the simulation step), since the `std::vector` implementation had to perform large number of dynamic memory allocations (hundreds or thousands for every simulation step). Then, we redesigned the index to be represented by two `std::vector`s of pre-determined sizes (the number of blocks for the first one and the number of particles for the other one) with data organized in a way similar to FAT filesystem: the first array contains one element for each block of space. It points to the second array to an offset that corresponds to the first particle that lies in the block. The second array contains one element for each particle. It stores an offset to the next particle in the list of particles that lie in the same block or an end-of-list flag. Together, these arrays form a linked list of particles for each block of space.

The effect of this optimization is quite interesting and points to a bigger and more important idea. But first, we have to point out that the index was used both on the host and on the Xeon Phi coprocessor. Both subsystems used exactly the same implementation, so both of them were affected by the optimization. On the host, the effect was not that significant (several percent, barely above measurement error). But on the Xeon Phi coprocessor, the effect was much greater. Depending on the setup, it could be as much as 40%. An important conclusion to draw from this observation is the following: the architecture of (multi-core) Xeon processors and Xeon Phi coprocessors is very similar, so the same optimization technique can usually be used for both systems.

But there are important differences between the designs of the two systems, so the same optimization may result in different performance gains when applied to both systems. The memory sub-system is a good example of such design difference and it could be a prime suspect in many of these situations. Both systems behave the same in the sense that they are fully cache-coherent, use virtual memory, and every core can access the whole memory range (if memory access rights permit), but the hardware implementation is very different. Efficient memory access and minimal memory sharing is important for the performance in both systems, but the effect is more pronounced on Xeon Phi coprocessors. There are other differences between the systems, like different vector units, but these are often abstracted by the auto-vectorization techniques of the compiler.

### B. Parallel processing

As we have already mentioned, the SPH is performed in four phases: preparatory phase, evaluation of gravity and density, evaluation of pressure, finalization (clean-up) phase. The first phase has not been parallelized, since there are complex data dependencies involved. The second and third phase are parallelized using our library. The last step is parallelized using just TBB, since the operation it performs is quite trivial and it would not benefit from Xeon Phi coprocessors, because the necessary data transfers would mitigate the benefits of having additional computing power.

From the theoretical point of view, the behavior of the second phase (the same is true for the third phase) is defined by a function $f(U, p)$, where $U$ is the whole problem state (the set of particles $P = \{p_1, \ldots, p_n\}$, the gravity field and spatial index), $p$ is a particle, and the result is also a particle. The whole phase transforms $U$ to $U'$ by transforming $P$ to $P' = \{f(U, p_1), \ldots, f(U, p_n)\}$. It is important to note, that the results of $f(U, p)$ only depend on the original problem state $U$, meaning that $f(U, p_i)$ can be evaluated independently of each other in parallel. To achieve this effect, the simulation had to be split into the aforementioned four phases.

So, after the `offload_for_each` (in phase two or three) starts, it transfers the whole $U$ to all of the coprocessors. Their copy of $U$ does not get updated, because after the offloaded for-loop finishes, it is not used any more. The host uses the coprocessors like this: it sends a particle $p$ to the coprocessor, lets it perform $p' := f(U, p)$ with the particle and the coprocessor's copy of $U$ and then retrieves the modified $p'$ so it can be used to build $U'$. Note that the particles are not sent as individual DMA transfers but they are grouped together to create larger blocks. This only affects scheduling and performance, not the result.

As we have already mentioned, one useful aspect of our library is the fact that it is possible to call `offload_for_each` multiple times in parallel or in parallel with other TBB tasks. We do that for rendering and simulation, because it is possible to render the result of N-th simulation step concurrently with the next simulation step, creating a two-step pipeline.

## VI. EXPERIMENTS

The experiments have been performed on a machine with two six-core Intel Xeon CPUs (X5680, 12M Cache, 3.33 GHz, Hyper-Threading) and two Xeon Phi coprocessors. The source codes were compiled with the Intel C++ Composer XE compiler set to O3 optimization level. All times are wall clock times, measured by the internal clock of the host machine. The stop watch was always started just before the simulation, after the initial setup. It was stopped right after the last frame got rendered.

### A. Configuration variants

In this section, we will use the same problem setup but different configuration of the run-time. We used one million particles and the video resolution was set to 100x100 pixels.

Table I shows the results. The first column shows the system configuration, the second tells whether simulation and rendering were executed in parallel (pipelined), the third gives the total processing time, the fourth shows the percentage of the simulation that was performed by the Xeon Phi coprocessors and the last column gives speedup over serial version. In our experiments we have used all available cores of the host, except for the serial execution where only one core of the host is used.

As you can see, there is a significant performance improvement when one card is used. The addition of the second card further improves the performance, although not by such a large margin. There are two reasons for this. First is the Amdahl's law. The preparatory and cleanup phases of each simulation step are done by the host. Second reason is the fact, that both cards are made to perform significant DMA transfers at about the same time. For example, at the beginning the whole problem set is transferred to all of the cards. Since the throughput of PCI Express is not infinite, this slows the whole execution down.

Another interesting result is the 2 to 8% speedup (depending on the configuration) that is achieved by pipelining the simulation and rendering steps. When the execution is pipelined, the rendering can use the cores that would otherwise be idle during the preparatory phase of the simulation. Even though the rendering takes longer than that, it can coexist with the parallel simulation phases, since they both use the TBB task scheduler. We know that this is happening, since the total wall clock time necessary to render the frames is larger if the execution is pipelined, so it means that some of the available resources have been allocated to the simulation while rendering was running. But the total time necessary to perform rendering and simulation in parallel is shorter than the time necessary to perform them one after another.

Table I
SPH PERFORMANCE FOR 1 000 000 PARTICLES USING VARIOUS CONFIGURATIONS OF THE HOST AND COPROCESSORS.

| system configuration | pipelined execution | total time [s] | work done by coproc. | speedup |
|---|---|---|---|---|
| serial (1 core) | no | 17174.40 | 0% | 1 |
| host (all cores) | yes | 1834.72 | 0% | 9.3 |
| host (all cores) | no | 1876.57 | 0% | 9.2 |
| host + 1 copr. | yes | 885.63 | 54% | 19.4 |
| host + 1 copr. | no | 927.31 | 55% | 18.5 |
| host + 2 copr. | yes | 628.13 | 71% | 27.3 |
| host + 2 copr. | no | 678.76 | 71% | 25.3 |

Further interesting figure is this: nearly 40% of instructions executed on the coprocessors were vector instructions. Considering that no specialized vector code was used in our implementation, this shows that the auto-vectorization done by the compiler is being used a lot.

The last observation is that the host only configuration (created by setting the number of coprocessors to zero) is in fact quite efficient way of building a parallel application. When compared directly to `parallel_for_each` from TBB, it can run several percent faster. This is thanks to the fact, that our solution is more specialized than the one in TBB, so we were able to cut some corners and significantly reduce overhead of our implementation. On the other hand `parallel_for_each` is more general and `parallel_for` (another parallel algorithm provided by TBB) would be faster for large number of items that are easy (measured by the number of instructions) to process.

### B. Different problem sizes

Figure 4 depicts SPH performance for various system configurations and numbers of particles. All variants use pipelined rendering, so the only difference in configuration is the number of coprocessors.

Figure 4(a) depicts execution time in seconds for each combination and Figure 4(b) depicts the same data, but as a speedup relative to the host-only variant. It is clear to see that with small data sizes, the overhead created by the communication with the coprocessors outweighs any possible performance gains. This is true for scenarios until around 50 thousand particles. The overhead consists of two parts: fixed overhead associated with initialization / clean-up and variable overhead associated with the necessity to transfer the problem state to the coprocessors. The size of the problem set is directly proportional to the number of particles. But, as the number of particles increases, so does the number of instructions that are necessary to process the particles. The computational requirements grow faster than the memory requirements, since the particles are more densely packed and interact more often with each other – interaction means computation. Both these effects stack up and as a result, larger data sets result in larger speedup gained by the use of Xeon Phi.

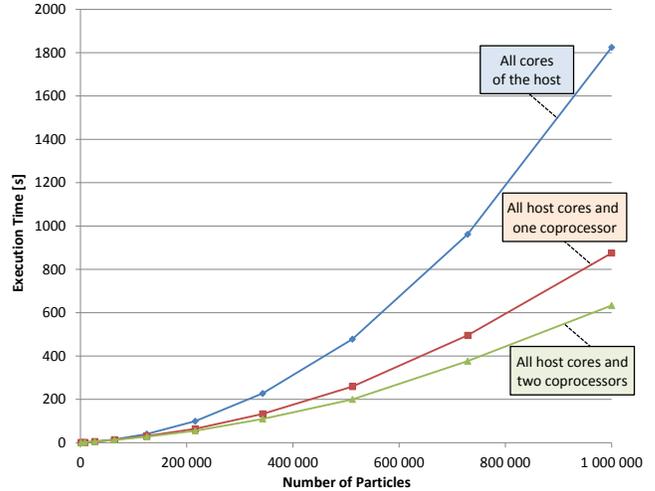

(a) Execution Time

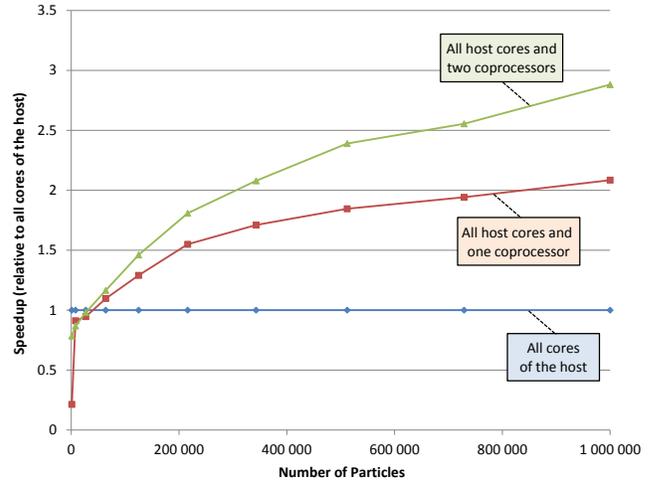

(b) Speedup

Figure 4. SPH performance for various system configurations and numbers of particles {1000, 8000, 27000, 64000, 125000, 216000, 343000, 512000, 729000, 1000000}.

## VII. RELATED WORK

Before the Xeon Phi architecture, there have been other experimental architectures that were based on similar ideas, most notably the Larrabee [5], which unlike Xeon Phi, was designed to be a GPU. Another example is the SCC architecture [6], which does not provide full cache coherency.

In the current range of tools that support Xeon Phi software development, two main groups can be identified. First, there is the SCIF library and other low-level tools. Second, there are solutions integrated into a compiler. There is the offloading support provided by the Intel C++ Composer XE compiler, which has already proved to be a good solution if hybrid execution is not desired [7]. The compiler's offloading support is usually combined with OpenMP to parallelize the execution after it was offloaded [8].

Naturally, Xeon Phi competes with graphic cards that are being used for general purpose computation (GPGPU) and the results certainly can be competitive [9]. CUDA and OpenCL, the best known development frameworks in GPGPU area, follow a different design to the one used on Xeon Phi but they also don't provide direct support for hybrid execution. It is possible to compile OpenCL codes for the host as well, so a similar solution to ours could probably be built. But it won't integrate as closely with the rest of the host-side process. For example, we significantly benefit from the fact, that there is one shared TBB thread pool for the whole application.

The C++ AMP [10] technology targets GPUs and, like our library, relies heavily on C++. However, the are major differences. The C++ AMP does not target hybrid execution and it requires compiler support. Still, it could probably be modified to support Xeon Phi and hybrid execution.

The serialization library that our project uses is our own creation, but there are other alternatives, like Boost Serialization [11], s11n [12], or Protocol Buffers [13]. They share many ideas with our solution, but our library is tuned for our purposes to minimize the overhead by minimizing the number of instructions necessary to serialize an object and also minimizing the size of the serialized object. A structure with two 4-byte fields is serialized as a sequence of 8 bytes, no extra identification or description is necessary, since both sides know at compile time what data is being transferred and how it is laid out.

## VIII. Conclusion

The goal of this paper was to show a new way of application development that has been made possible by the Xeon Phi coprocessors. The `offload_for_each` function template allows the developers to quickly build new applications that target the architecture. It is now sufficient to write just one source code and still create an application that can effectively use both the coprocessors and the resources of the host machine. The performance results obtained by our experiments support the viability of this approach.

We have also discovered that when built this way, the application is much easier to develop, debug and optimize, since it can easily be set up to run in a host-only configuration and then the whole execution can be monitored with tools that all developers are already familiar with.

In the future, we would like to extend the library to include more parallel algorithms (e.g., parallel reduction and pipeline). We will also investigate more sophisticated scheduling and work allocation strategies.

## Acknowledgment

This work was supported by the European Commission as part of the FP7 Project PEPPHER under grant 248481. We thank Intel Corporation for providing the system for our research. Intel, Xeon, and Xeon Phi are trademarks or registered trademarks of Intel Corporation or its subsidiaries in the United States and other countries. Other brands and names are the property of their respective owners.